# Are universal "anomalous" properties of glasses at low temperatures truly universal?

M. A. Ramos[1,2]

[1]Laboratorio de Bajas Temperaturas, Departamento de Física de la Materia Condensada,Universidad Autónoma de Madrid, Madrid, Spain
[2]Condensed Matter Physics Center (IFIMAC) and Instituto Nicolás Cabrera (INC), Universidad Autónoma de Madrid, Madrid, Spain

**Abstract** The specific heat $C_p$ and other properties of *glasses* (ranging from amorphous solids to disordered crystals) at low temperatures, are well known to be markedly different from those in fully-ordered crystals. For decades, this qualitative, and even quantitative, *universal* behavior of glasses has been thoroughly studied. However, a clear understanding of its origin and microscopic nature, needless to say a closed theory, is still lacking. To shed light on this matter, I review the situation in this work, mainly by compiling and discussing measured low-temperature $C_p$ data of many glasses and disordered crystals, as well as highlighting a few exceptions to that "universality rule". Thus, one can see that, in contrast to other low-temperature properties of glasses, the magnitude of the "glassy" $C_p$ excess at low temperature is far from being universal. Even worse, some molecular crystals without a clear sign of disorder exhibit linear coefficients in $C_p$ larger than those found in many amorphous solids, whereas a few of the latter show negligible values.

## 1 Introduction

Almost 50 years ago, Zeller and Pohl [1] demonstrated that low-temperature thermal properties of noncrystalline solids did not follow the expected behavior predicted by Debye theory, in clear contrast to insulating crystals. This fact was a bit striking, because long-wavelength acoustic vibrations dominating low-temperature thermal properties should be insensitive to atomic positional disorder [2, 3]. However, in all studied substances, also including earlier data from the literature, the measured specific heat $C_p$ of those amorphous solids or glasses below ~1 K exhibited [1] a linear dependence on $T$ instead of the purely cubic dependence observed in crystals and well explained by Debye theory. Moreover, in the cases where comparison was possible such as $SiO_2$, the low-temperature specific heat of the amorphous solid was found to be a few orders of magnitude larger than that of its crystalline counterpart. In addition, a broad maximum in $C_p/T^3$ currently known as the "boson peak" was ubiquitously observed at around 3–10 K in glasses [1, 3, 4], signalling a deviation from the expected horizontal level for a crystal at low enough temperatures. In fact, a corresponding broad peak in the reduced vibrational density of states over the frequency-squared Debye prediction for acoustic phonons, $g(\nu)/\nu^2$, has also systematically been observed in glasses by Raman or inelastic-neutron scattering [4–6].

In addition, the thermal conductivity $\kappa(T)$ of amorphous solids, or glasses in general, also looks very different to that in crystals [1, 3, 4]. Instead of the cubic increase with $T$ followed by a decrease due to phonon-phonon interactions typical of crystals, the thermal conductivity of the glass is orders of magnitude lower and increases quadratically with temperature, followed by a plateau, and then a further slow increase, in clear contrast to the crystal.

Those "anomalous" thermal properties found in amorphous solids at low temperatures soon were complemented by related findings in their acoustic and dielectric properties. Again, ultrasonic and dielectric experiments performed in amorphous solids showed a behavior completely different from that of crystalline solids [3]. For instance, the acoustic (and dielectric) absorption of glasses is strongly enhanced compared to crystals. At temperatures below ~100 K a broad Arrhenius-like absorption peak is usually observed, whereas at lower temperatures around 1 K a ubiquitous plateau in the associated internal friction $Q^{-1}$ dominates at keV frequencies followed by a dropoff at the lowest temperature, occurring at lower temperatures as the measuring frequency decreases [7].

After having identified the abovementioned *universal* behavior of amorphous solids (i.e. structural glasses) at low temperature in a number of them, it was natural in the 80's and 90's to start a search for glassy behavior in crystalline solids with some kind of disorder, beyond the translational disordered characteristic of noncrystalline (amorphous) solids. Firstly, alkali cyanide and other mixed crystals, which were grown with a controlled amount of orientational disorder leading to an orientationally-disordered state for appropriate concentrations, thus usually termed "orientational glasses", exhibited low-temperature specific heat and thermal conductivity very similar to those observed in structural (i.e. fully noncrystalline) glasses [7, 8]. Another type of very interesting "orientational glasses" which were studied later is that of so-called "glassy crystals" [9], that are produced by quenching plastic crystals and exhibit orientational disorder of dynamic origin within a cubic lattice of molecules. These glassy crystals of ethanol [10, 11] and other molecular solids [4] were found to present the same "glassy" behavior as genuine structural glasses. Hence it is more and more spoken about "universal low-temperature properties of glasses" than about those for amorphous solids, as it was usual at the beginning. Consequently, the origin of this "anomalous" behavior −in comparison to textbook crystals− is no longer ascribed to the lack of translational long-range order, but rather it tends to be related to some dynamic disorder inherently present in any non-fully-ordered solid.

Most recently, however, several "exceptions to the rule" have been reported: some crystals with a minimal amount of disorder also seem to exhibit glassy behavior at low temperature [12–14], whereas some genuine amorphous solids lack the linear-in-temperature contribution to the specific heat [15–17], which is the fingerprint of glassy anomalies.

The aim of this contribution within this special issue dedicated to M. A. Strzhemechny, who has been indeed very much interested in the low-temperature specific heat of molecular solids, is to review the still controversial and open question of the universal "anomalous" properties of glasses, mainly focused on their low-temperature specific heat.

In section 2, the phenomenology described above is extended by including some models and theories attempting to explain it, with special emphasis on the specific heat. A compilation and review of the main features of low-temperature specific-heat data of both structural glasses and disordered crystals is presented in section 3, followed by a brief discussion about to which extent this thermal property can be considered as universal. The conclusion after this data analysis and subsequent discussion is summarized in section 4.

**2 Low-temperature universal "anomalies" of glasses**

As described above, it is clear that low-temperature thermal properties of noncrystalline solids (i.e., amorphous solids, or glasses in general) differ remarkably from their crystalline counterparts. This "anomalous" (for unexpected) behavior is further considered as "universal" because: (i) any kind of noncrystalline substance (oxide glasses, amorphous thin films, polymers, organic molecular glasses, metallic glasses, even many disordered crystals…) exhibits these properties; (ii) some of these properties are even very similar quantitatively.

Simplifying, one can distinguish two distinct temperature ranges: $T < 1–2$ K, and $1–2$ K $< T < 10–20$ K, each with its different phenomenology.

## 2.1. $T < 1$–2 K

As already mentioned, the specific heat of glasses below 1–2 K ubiquitously exhibit a quasilinear dependence on temperature $C_\mathrm{p} \propto T$, hence decreasing with temperature much more slowly than in crystals that follow the cubic Debye law. In the same range, the thermal conductivity of glasses varies as $\kappa \propto T^2$ instead of cubically, but remains much lower than that of their crystalline counterparts. In addition, the internal friction $Q^{-1}$ is found to present a universal plateau $\sim 5 \times 10^{-4}$ with a dropoff at millikelvin temperatures. The corresponding sound velocity variation increases logarithmically with temperature in this lower temperature range. All these universal properties of glasses at $T < 1$–2 K [3, 4] are markedly different from those of canonical crystalline solids.

Most of these low-temperature properties soon were successfully accounted for by the Tunneling Model (TM). At least for genuine amorphous solids, the TM [18, 19] postulated a simple, random distribution of asymmetric double-well potentials arising from the configurational disorder inherent to solids lacking long-range translational order. Thus, additional low-frequency excitations (tunneling states or two-level systems, TLS) would appear in noncrystalline solids, ascribed to atoms or groups of atoms performing quantum tunneling motion between two configurations of similar potential energy. Basically with two simple parameters (a constant density $P_0$ of TLS per unit energy and volume, and a constant coupling energy $\gamma$ between phonons and TLS), the TM seemed able to rationalize even quantitatively the main glassy properties of glasses below 1–2 K [3], what justifies its wide recognition. In its simplest form, the density of TLS is independent of the splitting energy $n(E) = \mathrm{const} = n_\mathrm{TLS}$, and the specific heat is straightforwardly [18, 19] $C_\mathrm{p}(T) = (\pi^2/6)\, n_\mathrm{TLS}\, k_\mathrm{B}^2 T$.

Nevertheless, some doubts and criticisms about the TM has also been raised by several authors [20–23], who have argued how improbable it was that a random distribution of independent, *noninteracting* tunnelling two-level systems would produce essentially the same universal constant for the thermal conductivity or the acoustic attenuation at low temperatures, despite a wide distribution of material parameters among different substances. Also, some acoustic and dielectric experiments below 0.1 K have reported significant discrepancies with the TM for both metallic and insulating glasses [24–26].

## 2.2. 1–2 K $< T <$ 10–20 K

Above 1–2 K, where the glassy behavior is featured by the abovementioned boson peak and the plateau in thermal conductivity, the situation is even much more debated in the literature. Very different approaches and competing models have been proposed. For instance, Schirmacher [27] postulated a fluctuating elasticity theory, which essentially assumes a random distribution of elastic constants, to account for the transformation of Debye lattice dynamics in crystals into a vibrational density of states producing the boson peak in glasses. On the other hand, out of their Random First Order Theory (RFOT) of the glass transition, Lubchenko and Wolynes have associated the existence of two-level systems and the boson peak to cooperative motions of microscopic regions comprising a mosaic structure [28−30]. In a different view, other authors claim [31] that the boson peak is nothing else that of a smeared out van Hove singularity associated to transverse phonon-like vibrations in glasses.

Nonetheless, a very useful and relatively often employed approach to rationalize and fit experimental data of glasses at low temperature is provided by the Soft-Potential Model (SPM) [32, 33] and some derivations from it [34−36]. The SPM postulates the coexistence of Debye-like acoustic phonons with low-frequency quasilocalized anharmonic vibrations. These "soft modes" are related to a random distribution of quartic atomic potentials in glasses, which produces quasilocal configurations ranging from anharmonic double-well potentials (thereby including the TLS of the TM) to single-well potentials, more-or-less harmonic, which contain the vibrational modes responsible for the boson peak. Independently of the credit we ultimately give to the SPM, it is a very convenient and straight method to assess the different contributions to the $C_p$, and will be used in the following.

In brief, at low enough temperatures the specific heat of glasses follows the practical SPM equation [37]

$$C_p = C_{TLS}T + C_D T^3 + C_{sm} T^5 \qquad (1)$$

where $C_{TLS}$ is the linear coefficient ascribed to the TLS or tunneling states in agreement with the TM, $C_D$ is the usual Debye coefficient related to phonon-like acoustic vibrations, and $C_{sm}$ is the new contribution of the low-frequency *soft modes*, specifically those lying in the low-energy tail of the boson peak. Of course, this simplification in the low-temperature limit is only valid up to temperatures below the maximum in $C_p/T^3$. From eq. (1), it is straightforward that a simple least-squares quadratic fit within a plot $C_p/T$ vs $T^2$ directly provides the three sought coefficients. This fitting method for data analysis has been shown to be self-consistent (see Fig. 4 of Ref. [37]). Furthermore, the fifth-power coefficient associated with a vibrational density of states for "soft modes" following $g(\omega) \propto \omega^4$ at low frequency has been supported by recent studies [38−41].

Furthermore, following the procedure of eq. (1) based upon the SPM premises, an old open question was also unveiled. Along many years after the publication of the TM in 1972 [18, 19], a simple linear fit for $C_p/T$ : $T^2$ was routinely performed to determine the linear coefficient $C_{TLS}$ ascribed to the density of TLS, as well as the cubic coefficient ($C_3$). The latter coefficient was found to be systematically much larger than the Debye coefficient $C_{Debye}$, which can be directly obtained from the sound velocity and mass density of the material (see Table 3.1. in Ref. [3], or Table V in Ref. [42]). Hence it has been often stated that the "calorimetric" Debye coefficient of glasses is larger than the "elastic" one, i.e. $C_3 > C_{Debye}$. However, such a procedure clearly ignores the different contribution to $C_p$ related to the boson peak, which is not fully negligible at the temperatures of most of data fits [37].

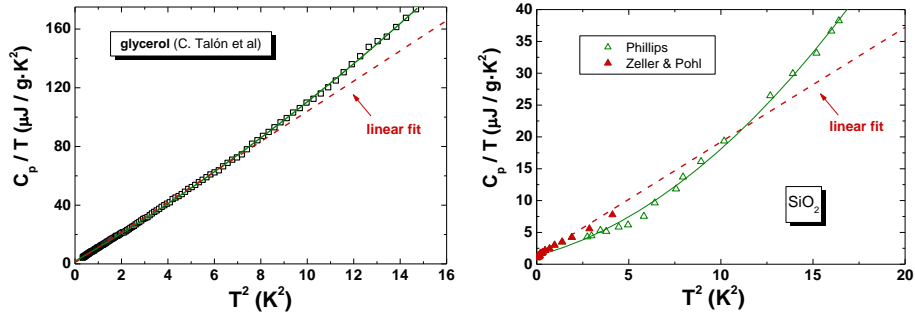

**Fig. 1** Specific-heat data of glycerol (left, after Ref. [43]) and of SiO$_2$ (right, after Refs. [1] and [5]), where a traditional linear fit (dashed lines) is compared to the proposed SPM quadratic fit of eq. (1) (solid lines). (Color figure online)

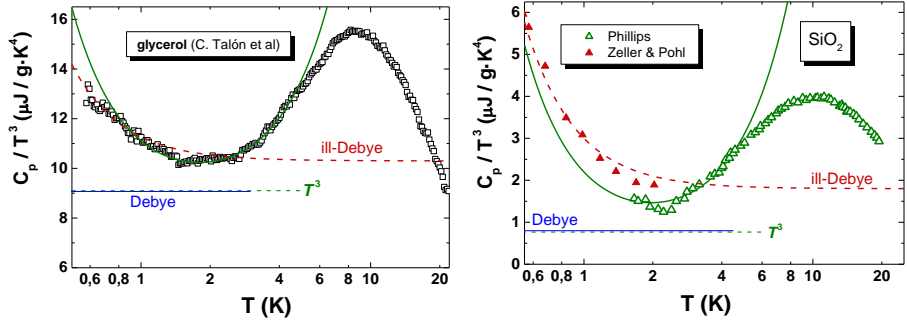

**Fig. 2** Same data of Fig. 1 in a much wider temperature range (in a log scale), using a Debye-reduced representation $C_p/T^3$ vs $T$. The traditional TM linear fits conducted in Fig. 1 are shown here to produce a wrong determination of the Debye level ("ill-Debye" dashed lines). On the contrary, the SPM quadratic fit (solid lines) imply a Debye level (horizontal dashed lines, labelled $T^3$) indistinguishable from the Debye coefficients obtained from elastic data (horizontal solid lines, labelled Debye). (Color figure online)

As shown in Fig. 1 (where the TM and SPM fits are compared) and in Fig. 2 (where the Debye-reduced specific heat $C_p/T^3$ is depicted) for both glycerol [43] and $SiO_2$ [1, 5] glasses, that linear fitting of $C_p/T : T^2$ plots gives an *ill-Debye* coefficient, which significantly overestimates the genuine cubic contribution of acoustic phonon-like vibrations in glasses. On the contrary, the SPM approach of eq. (1) correctly accounts for such contribution and provides a cubic coefficient $C_D$ in excellent agreement with the one obtained from elastic data (see Fig. 2 here, and Table 1 in Ref. [37]), i.e. $C_D = C_{Debye}$.

## 3 Review of specific heat data: Discussion

From all low-temperature properties exhibiting "anomalous" glassy behavior, this article focuses on the specific heat, that is probably the most relevant thermodynamic quantity, and the most abundantly measured property. As already said, the low-temperature specific heat of amorphous solids and other non-fully-crystalline solids exceeds that predicted on the basis of the Debye theory by a considerable amount [3, 4]. Specifically, $C_p$ is nearly linear in temperature below, say, 1 K, what is in principle well accounted for by the TM, as described above.

One striking aspect of this behavior is that the universality found is, to a large extent, also quantitative, though not so dramatically as the thermal conductivity or the acoustic attenuation. The linear coefficient $C_{TLS}$ ascribed to a density of TLS "defect modes" independent of energy has been stated to experience a modest variation from substance to substance on the order of 10–20 at most [20, 21].

Nonetheless, to compare $C_{TLS}$ per gram or per mole among different substances may be not too significative. Instead, we will scale the linear $C_{TLS}T$ contribution to the total specific heat $C_p$ evaluated at 1 K. Alternatively, we will assess the ratio $C_{TLS}T/C_{Debye}T^3$ also at 1 K, which amounts to the ratio of TLS to Debye coefficients $C_{TLS}/C_{Debye}$ expressed in $K^2$.

In Table 1 a compilation of both ratios is presented for a number of reported glasses, together with the temperature $T_{BP}$ at which their maximum in $C_P/T^3$ (boson peak) is observed. For a few of them, the linear coefficient $C_{TLS}$ was determined with data only above 1 K, what implies a less reliable evaluation of $C_{TLS}$. Theses cases are marked by an asterisk. There are two glasses included in the table where a dramatic depletion of TLS have been claimed, with $C_{TLS} = 0$ within experimental error: ultrastable indomethacin (IMC) [16] and toluene glass [17]. The latter was measured only down to 1.8 K and hence it may be more doubtful, but the former was measured well below 1 K and this conclusion is very much robust.

The abovementioned case of amorphous silicon [15], where an absence of TLS was reported, with no boson peak in $C_p/T^3$ nor a plateau in the thermal conductivity, is not included here for two reasons. First, it was reported later by the same group that the excess or not of specific heat relative to the Debye expectation crucially depended on the preparation conditions [48]. Second, there is no clear boson peak temperature to be considered for comparison.

Table 1. Linear TLS contribution to the specific heat scaled to the total specific heat (second column) or only to the Debye contribution (third column) for many different structural glasses (amorphous solids). The position of the boson peak maximum $T_{BP}$ is also indicated (fourth column). Materials marked with an asterisk signal that the linear coefficient $C_{TLS}$ was determined from data above 1 K, so being less reliable. They are displayed ordered by decreasing $T_{BP}$.

| MATERIAL | $C_{TLS}T/C_p$ (1 K) | $C_{TLS}T/C_{Debye}T^3$ (1 K) | $T_{BP}$(K) | Ref. |
|---|---|---|---|---|
| $(B_2O_3)_{0.75}(Na_2O)_{0.25}$ | 0.69 | 2.25 | 11 | 37 |
| $(B_2O_3)_{0.84}(Na_2O)_{0.16}$ | 0.61 | 1.6 | 10 | 37 |
| $SiO_2$ | 0.61 | 1.7 | 10 | 37 |
| glycerol | 0.16 | 0.20 | 8.7 | 37 |
| $GeO_2$ | 0.26 | 0.47 | 8 | 42 |
| $(B_2O_3)_{0.94}(Na_2O)_{0.06}$ | 0.49 | 1.0 | 7.5 | 37 |
| 1-propanol | 0.19 | 0.24 | 6.7 | 44 |
| H-ethanol | 0.43 | 0.77 | 6.1 | 44 |
| D-ethanol | 0.36 | 0.58 | 6.0 | 44 |
| $CaK(NO_3)_3$ | 0.42 | 0.73 | 6.0 | 37 |
| *n-butanol | 0.32 | 0.49 | 5.4 | 45 |
| $B_2O_3$ | 0.23 | 0.34 | 5.2 | 37 |
| PB | 0.22 | 0.30 | 5.1 | 37 |
| 2-propanol | 0.16 | 0.20 | 5.0 | 44 |
| *sec-butanol | 0.51 | 1.16 | 4.8 | 45 |
| *Isobutanol | 0.64 | 2.1 | 4.8 | 45 |
| *Toluene | 0 ± 0.3 | 0 ± 0.32 | 4.5 | 17 |
| PMMA | 0.22 | 0.30 | 3.6 | 37 |
| IMC (ultrastable) | 0 ± 0.02 | 0 ± 0.02 | 3.5 | 16 |
| IMC (conventional) | 0.22 | 0.28 | 3.5 | 16 |
| Amber (hyperaged) | 0.13 | 0.31 | 3.4 | 46 |
| Amber (rejuvenated) | 0.11 | 0.28 | 3.4 | 46 |
| Se | 0.026 | 0.033 | 3.1 | 37 |
| PS | 0.13 | 0.16 | 3.0 | 37 |
| Lexan | 0.085 | 0.13 | 2.7 | 42 |

Nevertheless, a linear contribution to the specific heat and a boson peak have also been reported in crystals with orientational disorder ("orientational glasses") [7, 8, 10, 11], but also in crystals with a minimal amount of disorder [12–14]. These are shown in Table 2, where the same ratios as in Table 1 are displayed for this distinct case of solids with glassy behavior.

Table 2. Linear TLS contribution to the specific heat scaled to the total specific heat (second column) or only to the Debye contribution (third column) for many different disordered crystals. The position of the boson peak maximum $T_{BP}$ is also indicated (fourth column). Materials marked with an asterisk signal that the linear coefficient $C_{TLS}$ was determined from data above 1 K, so being less reliable. They are displayed ordered by decreasing $T_{BP}$.

| MATERIAL | $C_{TLS}T/C_p$ (1 K) | $C_{TLS}T/C_{Debye}T^3$ (1 K) | $T_{BP}$(K) | Ref. |
|---|---|---|---|---|
| *CCl$_4$ | 0.70 | 2.5 | 9.2 | 14 |
| *CBrCl$_3$ | 0.68 | 2.3 | 7.7 | 14 |
| *CBr$_2$Cl$_2$ | 0.59 | 1.6 | 7.5 | 14 |
| 2-BrBP | 0.042 | 0.044 | 7.2 | 13 |
| H-ethanol-OG | 0.46 | 0.88 | 6.8 | 44 |
| D-ethanol-OG | 0.39 | 0.66 | 6.4 | 44 |
| *Freon-113 | 0.29 | 0.47 | 5.0 | 47 |
| PCNB | 0.13 | 0.16 | 4.8 | 12 |
| *Freon-112 | 0.33 | 0.56 | 4.5 | 47 |

Interestingly, and beyond those few exceptions to the rule already indicated, if we inspect the $C_{TLS}T/C_p$ ratios in Table 1 (comprising data from 25 structural glasses) and even in Table 2 (for differently disordered crystals), we can observe that the relative contribution of $C_{TLS}T$ to the total $C_p$ at 1 K (i.e. the relative density of TLS) is not so universal. Even better, if we look at the $C_{TLS}/C_{Debye}$ ratio, which is devoid of the double contribution of TLS to the numerator and the denominator, one finds almost two orders of magnitude of spread!

In order to search for some kind of trend or correlation, the obtained $C_{TLS}T/C_p$ ratios have been plotted in Fig. 3 versus the boson peak temperature, for both structural glasses (open squares) and disordered crystals (solid circles). Estimated error bars for those dimensionless ratios are also included for structural glasses. They are obtained from the statistical errors of the fits when data are from this author, and are just reasonable estimations from the reported coefficients by other authors in the literature.

In Fig. 4, the same specific-heat data are presented, but now showing the $C_{TLS}T/C_{Debye}T^3$ ratio, hence properly scaling the TLS contribution to the corresponding elastic one.

In both Fig.3 and Fig.4, the scaled $C_{TLS}$ magnitude seems to increase with increasing $T_{BP}$, as suggested by the dashed-dotted eye-guide line in Fig.4, though with some clear deviations from that general trend.

Another interesting observation from Fig. 3 concerns the two glasses with reported null $C_{TLS}$ coefficient, and hence without TLS, as pointed out above. Now it becomes clearer that the claimed depletion of TLS in ultrastable glasses of IMC is robust, whereas the case of toluene entails an error bar comparable to values in other glasses, which do exhibit a nonzero linear coefficient, such as selenium.

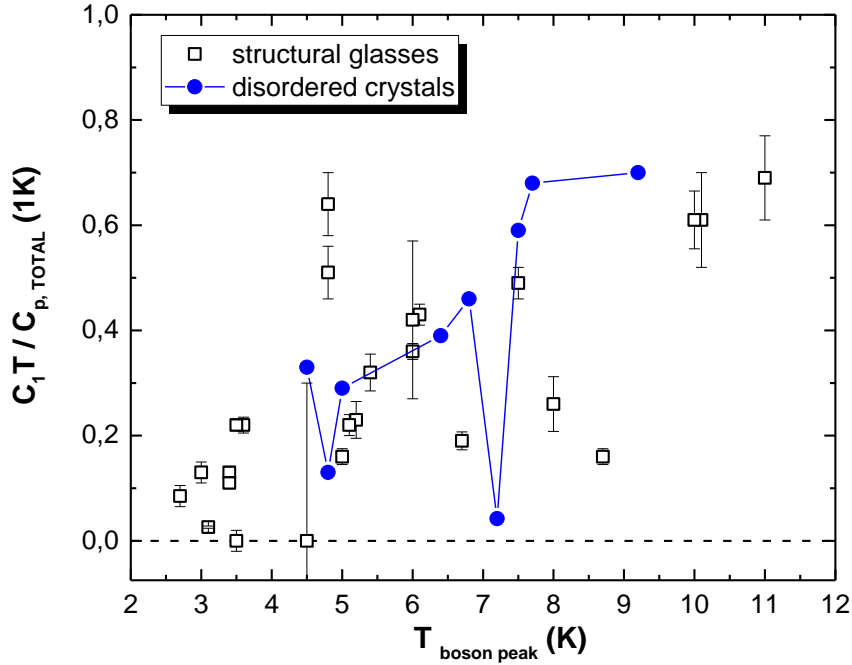

**Fig. 3** Relative fraction of the linear specific heat ascribed to TLS from the total specific heat of the material evaluated at 1 K, for different glasses (open squares) and disordered crystals (solid circles), listed in Table 1 and Table 2, respectively. (Color figure online)

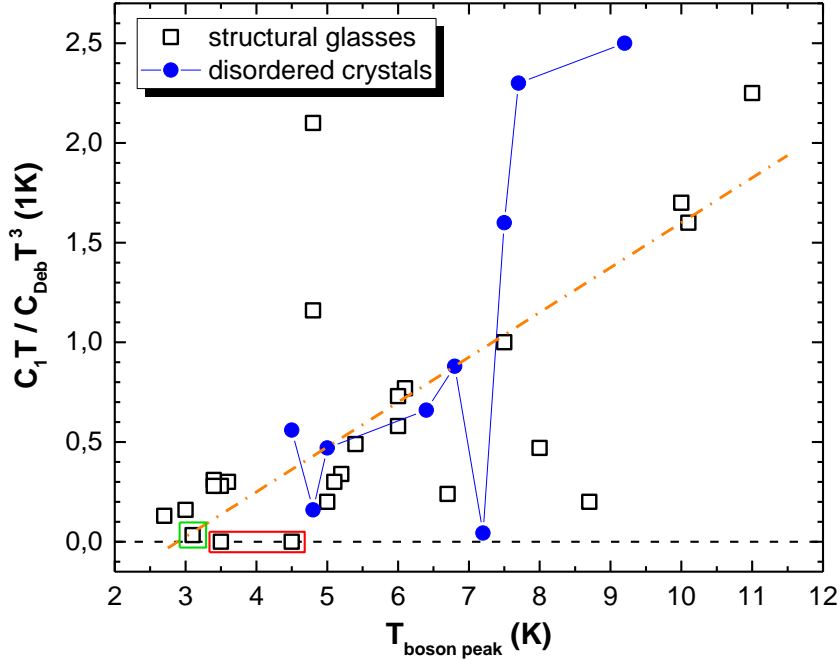

**Fig. 4** Ratio of the linear specific heat ascribed to TLS relative to the Debye contribution evaluated at 1 K, for different glasses (open squares) and disordered crystals (solid circles), listed in Table 1 and Table 2, respectively. The datapoint embraced by a green square corresponds to Se and those embraced by a red rectangle correspond to ultrastable IMC and toluene glasses. Dashed-dotted line is a guide for the eye. (Color figure online)

## 4 Conclusion

In summary, in this article I have reviewed measurements of the specific heat at low temperature in many glasses and disordered crystals. It has been shown that the apparently well-known universality of "glassy behavior" at low temperatures ascribed to a comparable amount of density of TLS ($n_{TLS}$) is far from clear. First, several exceptions to this universal behavior have been found (glasses with essentially null linear term in $C_p$, whereas some crystals with a minimal amount of disorder do exhibit such linear contribution). Second, when properly scaled, the dispersion in $n_{TLS}$ (i.e., a wide spread in the scaled linear coefficient $C_{TLS}/C_{Debye}$) is relatively large. Furthermore, this cast doubts on some reported absence of TLS in a few structural glasses (including our own results). Whereas in cases as ultrastable glass of IMC experimental error bars are extremely low and a dramatic depletion of TLS seems a robust finding, in other cases as toluene (measured at not so low temperatures) the upper bound is comparable to more accurate (small) values of the reduced $C_{TLS}/C_{Debye}$, such as that for amorphous Se.

All in all, $n_{TLS}$ could vary orders of magnitude (in proportion to the lattice vibrations contribution) among different glass-forming substances, which seems something, in principle, more reasonable or expected. The enigma would remain why this translates into much more universal values in glasses for the thermal conductivity at low temperatures, the plateau in the internal friction Q-1, the sound-velocity logarithmic shift with temperature, etc. In other words, why the so-called tunneling strength $C \equiv \gamma^2 P/\rho v^2$ varies so little despite a much larger fluctuation in any of these four material parameters.

**Acknowledgements** I am deeply grateful to Anthony J. Leggett and Uli Buchenau for encouraging discussions about this topic. Financial support from the Spanish Ministry of Science through project FIS2017-84330-R, as well as from the Autonomous Community of Madrid through program S2018/NMT-4321 (NANOMAGCOST-CM), is also acknowledged.